\documentclass[times,twocolumn,final]{elsarticle}
\usepackage{medima}
\usepackage{framed,multirow}

\usepackage{amssymb}
\usepackage{latexsym}
\usepackage{amsmath}
\usepackage{float}
\usepackage{setspace}

\usepackage{algorithm}
\usepackage{algorithmicx,algpseudocode}
\usepackage{booktabs}
\usepackage{array}
\newcolumntype{P}[1]{>{\centering\arraybackslash}p{#1}}
\newcolumntype{L}[1]{>{\raggedright\let\newline\\\arraybackslash\hspace{0pt}}m{#1}}
\newcolumntype{C}[1]{>{\centering\let\newline\\\arraybackslash\hspace{0pt}}m{#1}}
\newcolumntype{R}[1]{>{\raggedleft\let\newline\\\arraybackslash\hspace{0pt}}m{#1}}

\usepackage{url}
\usepackage{xcolor}

\usepackage{hyperref}

\definecolor{newcolor}{rgb}{.8,.349,.1}

\begin{document}

%\verso{Given-name Surname \textit{et~al.}}

\begin{frontmatter}

\title{Fast and accurate sparse-view CBCT reconstruction using meta-learned neural attenuation field and hash-encoding regularization}%

%\tnotetext[tnote1]{This is an example for title footnote coding.}

\author[1]{Heejun \snm{Shin}}

\author[1]{Taehee \snm{Kim}}
%\author[2]{Given-name3 \snm{Surname3}}

%% Third author's email
% \ead{author3@author.com}
\author[2]{Jongho \snm{Lee}}
\author[3]{Se Young \snm{Chun}}
\author[4]{Seungryong \snm{Cho}}
\author[1]{Dongmyung \snm{Shin}\corref{cor1}}
\cortext[cor1]{Corresponding author. 
 email: shinsae11@radisentech.com;}

\address[1]{Artificial Intelligence Engineering Division, Radisen Co. Ltd., Seoul, Republic of Korea}
\address[2]{Laboratory for Imaging Science and Technology, Department of Electrical and Computer Engineering, Seoul National University, Seoul, Republic of Korea}
\address[3]{Intelligent Computational Imaging Laboratory, Department of Electrical and Computer Engineering, Seoul National University, Seoul, Republic of Korea}
\address[4]{Medical Imaging and Radiotherapy Laboratory, Department of Nuclear and Quantum Engineering, Korea Advanced Institute of Science and Technology, Daejean, Republic of Korea}

%\received{1 May 2013}
%\finalform{10 May 2013}
%\accepted{13 May 2013}
%\availableonline{15 May 2013}
%\communicated{S. Sarkar}

\begin{abstract}
%%%
Cone beam computed tomography (CBCT) is an emerging medical imaging technique to visualize the internal anatomical structures of patients. During a CBCT scan, several projection images of different angles or views are collectively utilized to reconstruct a tomographic image. However, reducing the number of projections in a CBCT scan while preserving the quality of a reconstructed image is challenging due to the nature of an ill-posed inverse problem. Recently, a neural attenuation field (NAF) method was proposed by adopting a neural radiance field algorithm as a new way for CBCT reconstruction, demonstrating fast and promising results using only 50 views. However, decreasing the number of projections is still preferable to reduce potential radiation exposure, and a faster reconstruction time is required considering a typical scan time. In this work, we propose a fast and accurate sparse-view CBCT reconstruction (FACT) method to provide better reconstruction quality and faster optimization speed in the minimal number of view acquisitions ($<$ 50 views). In the FACT method, we meta-trained a neural network and a hash-encoder using a few scans (= 15), and a new regularization technique is utilized to reconstruct the details of an anatomical structure. In conclusion, we have shown that the FACT method produced better, and faster reconstruction results over the other conventional algorithms based on CBCT scans of different body parts (chest, head, and abdomen) and CT vendors (Siemens, Phillips, and GE).
%%%%
\end{abstract}


\end{frontmatter}

%\linenumbers

\section{Introduction}
\label{sec1}
Cone beam computed tomography (CBCT) is an emerging medical imaging technique to visualize the internal anatomical structures of patients. Compared to a fan beam CT (FBCT), CBCT takes advantages of higher image resolution and faster scanning time \citep{lechuga2016cone}. During a CBCT scan, a cone-shaped X-ray beam is diverged from a source which circularly rotates an anatomy of interest with an X-ray detector. This results in several projection images of different angles (i.e., views) that are collectively utilized to reconstruct a tomographic image. Nowadays, CBCT has been widely adopted as a practical diagnosis tool in human anatomies, including teeth \citep{scarfe2006clinical}, extremities \citep{carrino2014dedicated}, and chest \citep{hohenforst2014cone}. 

However, the amount of ionizing radiation to a patient during a CBCT scan is much higher than in conventional radiography, preventing its wide applications \citep{lechuga2016cone}, \citep{kan2008radiation}. Reducing the number of projections (i.e., sparse-view; $<$ 100 views) in a CBCT scan can effectively circumvent the high exposure; however, preserving the quality of a reconstructed image is challenging due to the nature of an ill-posed inverse problem \citep{gao2014low}.

Conventional CBCT reconstruction methods such as the Feldkamp-Davis-Kress (FDK) algorithm \citep{feldkamp1984practical} provide best image quality in an ordinary CBCT scan with hundreds of views but suffering from streak artifacts in a sparse-view scan. Iterative optimization algorithms, such as simultaneous algebraic reconstruction technique (SART) or adaptive steepest descent-projection on convex subsets (ASD-POCS) \citep{andersen1984simultaneous, sidky2008image}, produce good image quality in a limited number of views, but they require more computational time and produce unsatisfactory results as the number of views becomes very limited (e.g., 50 views or less).

As more recent approaches, several deep learning-based CT reconstruction techniques, primarily based on convolutional neural networks (CNNs) have successfully promoted superior image quality in a sparse-view setting. Those methods are typically categorized as one of three: sinogram-domain methods \citep{lee2018deep, anirudh2018lose, tang2019projection}, that directly interpolate or extrapolate sinograms to fill out missing information, images-domain methods \citep{jin2017deep, zhang2018sparse, georgescu2020convolutional, liu2020tomogan} which use CNNs to recover reconstructed images with artifacts, and dual-domain methods \citep{hu2020hybrid, wang2021improving, lin2019dudonet, zhou2022dudodr, wu2021drone} that combine these two approaches to utilize mutual information in both domains. However, those methods, which rely on each sinogram to reconstruct each 2D CT slice, are limited to being generalized to CBCT reconstruction, which requires using multiple 2D projections simultaneously to reconstruct a 3D tomography. In addition, the methods inherently require a lot of training CT data and suffer from potential training biases.

Recently, a neural radiance field (NeRF) algorithm \citep{mildenhall2021nerf} has been gaining tremendous popularity as an effective way to synthesize novel views of natural scenes or objects using a set of captured views. In NeRF, rather than a complex CNN, a simple feed-forward network is adopted to learn the relationship (i.e., implicit neural representation \citep{park2019deepsdf}) between the spatial coordinates in a 3D volume and metrics for these coordinates, such as colors and densities, to synthesize novel views. Based on the NeRF method, \cite{zha2022naf} proposed a new self-supervised method for CBCT reconstruction, which is called a neural attenuation field (NAF), demonstrating fast CBCT reconstruction with good quality using only 50 projections. Identical to the NeRF method, the NAF method does not require any training data and produces better reconstruction results than other conventional methods.

Although the NAF method reported promising results as a novel way for CBCT reconstruction, some room for improvement remains. First, a faster reconstruction time is more favorable since the convergence time of NAF is still relatively long compared to a typical CBCT scan time  \citep{scarfe2008cone}. Second, decreasing the number of projections, even less than 50 views, is more preferable, unless the image quality is acceptable for certain clinical purposes (e.g., patient positioning \citep{thilmann2006correction}, emergency scan \citep{jacques2021impact}, etc.), to reduce the amount of radiation exposure.

In this work, we propose a fast and accurate sparse-view CBCT reconstruction (FACT) method, which utilizes a meta-learning framework and a novel feature-encoding technique to alleviate potential disadvantages of NAF. In the FACT method, we first pre-trained a neural network using a small set of CBCT scans (= 15 scans) based on a meta-learning framework. Then, this meta-learned network is utilized to reconstruct a 3D tomography quickly using a new CBCT scan at test time. During the reconstruction, we regularized the hash-encoding process to further optimize the details of the anatomical structure even in the minimal number of views ($<$ 50 views).

In summary, the main contributions of this paper are the following:

-	We propose a FACT method that provides faster optimization speed and better CBCT reconstruction quality than a NAF method in the minimal number of views ($<$ 50 views) based on a meta-learning framework and a novel hash-encoding regularization process.

-	Unlike previous supervised CT reconstruction methods that extensively require a large amount of CT data ($>$ a few hundred) to train neural networks, the FACT method requires only a small number of scans (= 15 scans) to meta-train a neural network, avoiding preparation of a large amount of data and potential biases in AI training.

-	Throughout the extensive experiments, we have shown that, regardless of CT vendors (Siemens, Phillips, and GE) and body parts (chest, head, and abdomen), the FACT method produced consistently better CBCT reconstruction results than the NAF method.

\section{Related Works}
\subsection{Sparse-View CT Reconstruction}
Classical CBCT reconstruction algorithms \citep{feldkamp1984practical, andersen1984simultaneous, sidky2008image} include analytical optimization methods (e.g., FDK) that estimate attenuation coefficients based on the inverse Radon transform and iterative methods (e.g., SART, ASD-POCS) that apply iterative reconstruction algorithms until convergence. However, FDK works best when hundreds of projections are available and suffers from artifacts in a sparse-view scan. In contrast, iterative methods, such as ASD-POCS, can suppress artifacts but consume more memory and computational time.

In recent years, many deep learning-based approaches have been proposed for sparse-view CT reconstruction. Those works can be categorized into three methods: sinogram-domain, image-domain, and dual domain. Sinogram-domain methods enhance projection data (i.e., sinogram) using neural networks prior to image reconstruction. \cite{lee2018deep} proposed an Unet-based network to fill out the missing parts in sparsely-sampled sinograms, and \cite{tang2019projection} applied a CNN-based super-resolution method to up-sample low-resolution sinograms. On the other hand, image-domain methods improve the quality of reconstructed images using neural networks. \cite{jin2017deep} trained an Unet using reconstructed images from sub- and full-sampled sinograms to restore input CT images with artifacts. \cite{zhang2018sparse} improved the quality of reconstructed images using a DenseNet and a deconvolution-based network. Finally, dual-domain methods combine information in both domains for reconstruction. \cite{lin2019dudonet} proposed a novel radon inversion layer that concurrently refined sinograms and corresponding images. Similarly, \cite{zhou2022dudodr} introduced a new network layer called an image and projection data consistency layer. However, most of the works above utilized CNNs with a lot of training data, and those methods are challenging to be generalized to CBCT reconstruction.

More recently, with the success of an implicit neural representation for a novel view synthesis \citep{mildenhall2021nerf}, some studies \citep{zang2021intratomo, zha2022naf} adopted a NeRF algorithm for CBCT reconstruction and proposed self-supervised methods that parameterized attenuation coefficients as continuous attenuation field, reporting promising results using only tens of projection images.

%%%%%%%%%%%%%%%%%%%%%%%%%%%%%%%%%%%%%%%%%%%%%%%%%%%%%%%%%%%

\begin{figure*}[th]
\centering
\includegraphics[scale=1.0]{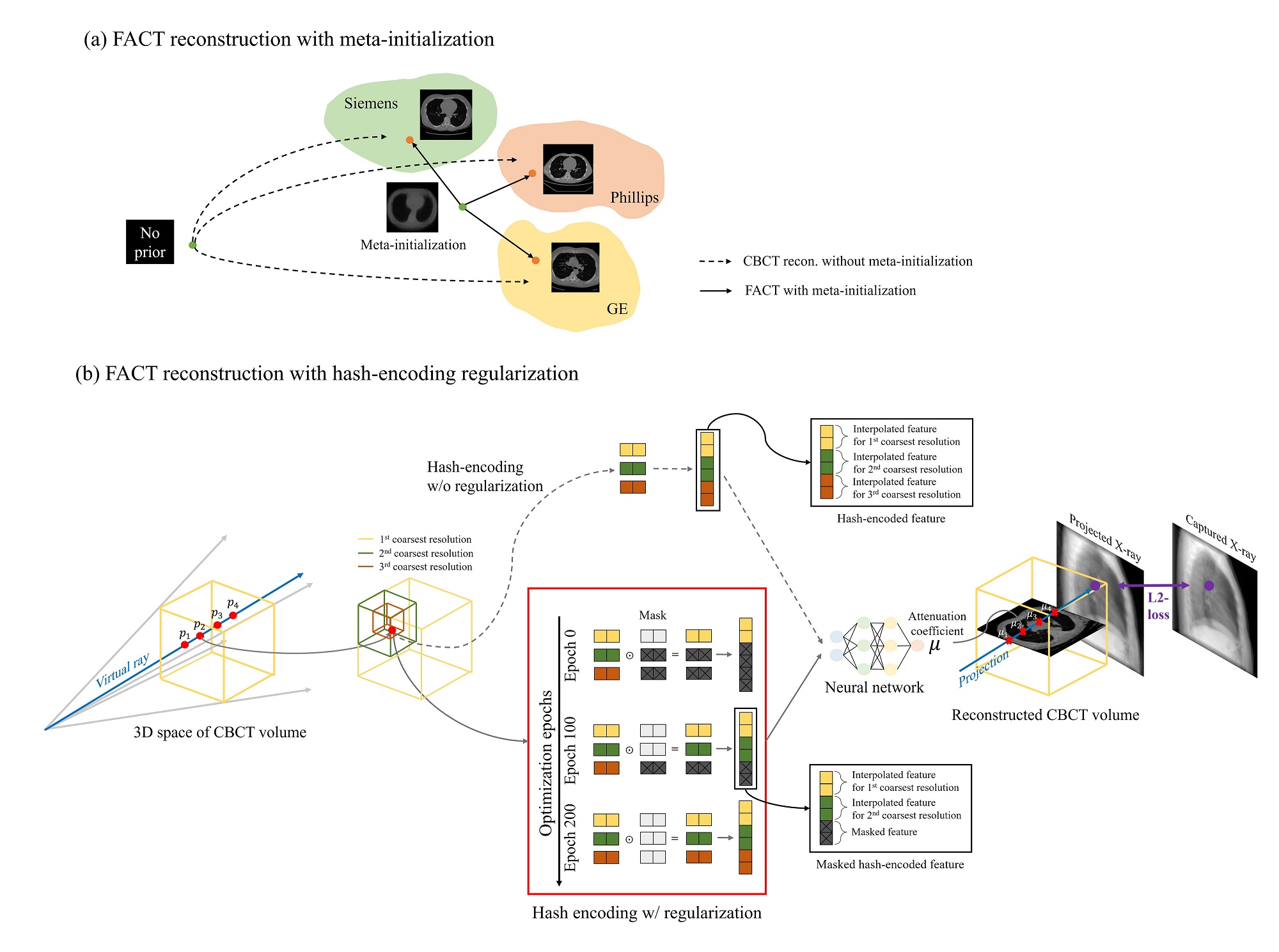}
\caption{Illustration of the FACT reconstruction with the meta-initialization and hash-encoding regularization. (a) By meta-learning the CBCT reconstruction functions (i.e., meta-initialization), the FACT reconstruction quickly optimizes each target scan to a specific CBCT image (dotted line vs. solid line). (b) In the FACT reconstruction, the hash-encoder, which includes feature vectors of multiresolution 3D grid points, generates feature embeddings (hash-encoded feature) of the query points (e.g., red point $p_2$) using tri-linear interpolation for each resolution (interpolated features). In the FACT method, a novel regularization technique (solid line and red box) for the hash-encoder is incorporated, which masks out the interpolated features depending on the optimization epochs, resulting in the masked hash-encoded feature.}
\label{fig:1}
\end{figure*}

%%%%%%%%%%%%%%%%%%%%%%%%%%%%%%%%%%%%%%%%%%%%%%%%%%%%%%%%%%%

\subsection{Neural Radiance Field}
NeRF \citep{mildenhall2021nerf} utilizes a neural network to map 3D spatial coordinates to values, such as colors and densities, using a set of 2D views. Despite the success of the original NeRF, it requires many input views and a lengthy optimization time. To solve these problems, \cite{yu2021pixelnerf} and \cite{chen2021mvsnerf} adopted CNNs to speed up a neural encoding process in NeRF. \cite{tancik2021learned} used a meta-learning framework to pre-initialize a neural network in NeRF, and \cite{jain2021putting} introduced auxiliary semantic consistency loss to supervise a neural network to synthesize a new scene with a limited number of input views. Other studies, such as \cite{niemeyer2022regnerf, yang2023freenerf}, proposed novel regularization techniques to reduce the number of views effectively.

In NeRF, positional encoding is commonly utilized to encode low-dimensional input features for a neural network to high-dimensional features to prevent the bias of the network toward learning low frequency components in the input images \citep{rahaman2019spectral, ronen2019convergence}. This encoding process helps the neural network map high-frequency components for a novel view generation \citep{mildenhall2021nerf, tancik2020fourier}. One of the popular encoding approaches is to discretize a 3D space into multiresolution grids and interpolate trainable features in different resolutions. \cite{takikawa2021neural} utilized a tree structure to save and manage multi-scale features in a 3D grid. \cite{muller2022instant} adopted a hash table to efficiently control the trainable features' memory while producing small and enriched features, making the optimization process faster.

\subsection{Optimization-Based Meta-Learning}
Optimization-based meta-learning aims to train a pre-initialized AI algorithm that can be rapidly adopted to unseen tasks using a few examples. This method usually utilizes two models, which are an inner model (i.e., task-specific learner) and an outer model (i.e., meta-learner), to find the best initialization of the outer model that will be optimized fast within a few gradient steps. Modal-agnostic meta-learning (MAML) \citep{finn2017model} first updates an inner model by sampling a task from a set of multiple tasks and gradually updates the outer model based on the updated weights of the inner model. For the faster adaptation of an AI model, \cite{lee2018gradient} incorporated task-specific information into MAML by combining meta-learned activation layers in the inner model. Similarly, \cite{flennerhag2019meta} applied a warped gradient descent optimization to the inner model. Other studies, such as \citep{nichol2018first, rajeswaran2019meta} tried to simplify the optimization process of MAML for faster adaptation. For example, a Reptile method \citep{nichol2018first} approximated MAML using first-order gradient optimization to reduce the computational time and memory burdens.

\section{Materials and Methods}

\subsection{CBCT Reconstruction Based on Implicit Neural Representation} \label{section:3.1}
The goal of CBCT reconstruction is to extract 3D tomographic information using a set of 2D X-ray projections. Compared to a FBCT scan, a CBCT scan uses a circularly rotating X-ray source with a panel detector to acquire a set of \(N\) 2D X-ray images with different views (i.e., X-ray projection set; \( \bold{X} = \{X_1, X_2, \ldots, X_N \in \mathbb{R}^{H \times W}\} \), where \(H\) and \(W\) are the height and width of images, respectively). The CBCT acquisition (\( \varphi \)), therefore, depends on a 3D object of interest (\(C\)), scanning angles of each view \( \left( \boldsymbol{\alpha} = \{\alpha_1, \alpha_2, \ldots, \alpha_N\} \right) \), and other parameters (\( \beta \); e.g., distance between X-ray source and detector, detector resolution, etc.).

\begin{equation}
\varphi(C, \boldsymbol{\alpha},\beta) \rightarrow \bold{X}
\end{equation}

Then we can formulate the CBCT reconstruction problem to find an inverse function \(F\) that accurately estimates an attenuation coefficient (\( \mu \in R \)) on a query point (\( p \in R^{3} \)) in a 3D object (i.e., \(\mu = C(p)\))  using the acquired set of 2D X-ray images (\(\bold{X}\)) and other parameters (\(\boldsymbol{\alpha}\) and \(\beta\)):

\begin{equation}
\mu = C(p) = F(p;\bold{X},\boldsymbol{\alpha}, \beta)
\end{equation}

To achieve this goal, conventional approaches \citep{feldkamp1984practical, andersen1984simultaneous, sidky2008image}, such as analytical methods (e.g., back-projection, etc.), recover attenuation coefficients in the uniform grid points of a 3D volume. In contrast, our approach, similar to \cite{zha2022naf}, introduces a mapping function (\(f_\theta\)), which is a feed-forward neural network, to map a continuous spatial coordinate inside the 3D volume to a corresponding attenuation coefficient. In other words, the neural network was trained to implicitly learn the spatial relationship between the coordinates and attenuation coefficients (i.e., implicit neural representation \citep{park2019deepsdf}). Before feeding the 3D coordinate into the network, we encoded it using a trainable hash-encoder (\(h_\phi\)) \citep{muller2022instant}:

\begin{equation}
f_{\theta}(h_{\phi}(p)) \rightarrow \mu
\end{equation}

To train those two functions (\(f_\theta\) and \(h_\phi\)), a virtual ray was cast from a cone beam source to a 3D space, passing through M number of coordinates (\(p_1,p_2,\ldots,p_M\)) along the ray trajectory and, consequently, producing \(M\) number of attenuation coefficients (\(\mu_1,\mu_2,\ldots,\mu_M\)) from the hash encoder (\(h_\phi\)) and neural network (\(f_\theta\); see Fig.~\ref{fig:1} b). Then, we can project another ray in the same direction of the ray casting and calculate the intensity of a point after the projection (e.g., purple point on a projected X-ray image in Fig.~\ref{fig:1} b) based on Beer's Law \citep{zha2022naf}:

\begin{equation}
I=I_0\,exp \left( -\sum^{M}_{i=1}\mu_i\delta_i \right)=I_0\,exp \left(  -\sum^M_{i=1}f_{\theta}(h_{\phi}(p_i))\delta_i \right)
\end{equation}

where $I$ is an X-ray intensity after the projection, $I_0$ is an initial intensity, and $\delta$ is the distance between the two adjacent coordinates. Assuming that $\tilde{I}$ is a point intensity in an acquired 2D X-ray image and matched with the projected point, then the two mapping functions (i.e., $f_\theta$ and $h_\phi$) can be optimized by minimizing the difference in the intensities between these two points using backpropagation. We can perform the optimization by iterating all the acquired X-ray images (\(\bold{X} = \{X_1, X_2, \ldots, X_N \} \)), considering all the points in each image (i.e., $H\times W$ number of points), and minimizing a mean squared error as follows:

\begin{equation}
argmin_{\theta, \phi} L(\theta, \phi, \bold{X}) = argmin_{\theta, \phi} \sum^N_{k=1}L'(\theta, \phi, X_k)
\end{equation}

$$
where \; L'(\theta, \phi, X_k) = \frac{1}{H \times W} \sum^{H\times W}_{j=1}(\tilde{I}_{j,k}-I_{j,k})^2
$$

$$
= \frac{1}{H \times W} \sum^{H \times W}_{j=1} \left( \tilde{I}_{j,k} -  I_0 exp\left(  
-\sum^M_{i=1}f_{\theta}(h_{\phi}(p_{i,j,k}))\delta_{i,j,k}  \right)    \right)^2
$$

Later, the optimized functions are used to fill out the attenuation coefficient values inside a 3D volume for CBCT reconstruction.

Here, the $f_\theta$ and $h_\phi$ are initialized with random weights, and, therefore, those functions should be optimized from scratch during a test time for a single CBCT data, requiring much computational time. To address this issue, we propose a new strategy to pre-train those two functions using a small set of CBCT scans and regularize them to improve the optimization speed and reconstruction quality.

\subsection{FACT Method}
\subsubsection{FACT Reconstruction with Meta-Initialization} \label{section:3.2.1}
In the FACT method, a meta-learning framework was adopted to initialize learnable weights of the mapping functions, $f_\theta$ and $h_\phi$ (see Section \ref{section:3.1}), using a few CBCT scans (15 scans; see Section \ref{section:3.3}). For this purpose, we modified a Reptile method \citep{nichol2018first} to update and initialize those two functions by utilizing sets of X-ray projections (i.e., no reconstructed CBCT images are required) for meta-initialization. Pseudo code for CBCT meta-initialization is shown in Algorithm \ref{algorithm:1}.

\begin{algorithm}
\caption{CBCT Meta-Initialization}
\begin{algorithmic}
\renewcommand{\itemsep}{0.5em}
    \Require Prepare $P$ X-ray projection sets $\{\bold{X_1}, \bold{X_2}, \ldots, \bold{X_P} \}$
    \State Randomly initialize weights $\theta_0$ of $f_\theta$ and $\phi_0$ of $h_\phi$
    \For{iteration $= 1, 2, \ldots$}
        \State Randomly sample a projection set (\(\bold{X}\))
        \State \(\theta'_{0} = \theta_0\)
        \State \(\phi'_{0} = \phi_0\)
        \For{epoch $= 1, 2, \ldots , K$}
            \State Update $\theta'_0=\theta'_0 - \gamma \nabla_{\theta'_0}L(\theta'_0 , \phi'_0, \bold(X))$
            \State Update $\phi'_0=\phi'_0 - \gamma \nabla_{\phi'_0}L(\theta'_0 , \phi'_0, \bold(X))$
        \EndFor
        \State Update $\theta_0=\theta_0 - \epsilon(\theta'_0 - \theta_0)$
        \State Update $\phi_0=\phi_0 - \epsilon(\phi'_0 - \phi_0)$
    \EndFor

    \State $\theta^*_0 \leftarrow \theta_0$
    \State $\phi^*_0 \leftarrow \phi_0$
\end{algorithmic}
\label{algorithm:1}
\end{algorithm}

Using the $P$ X-ray projection sets (\(\bold{X_1},\bold{X_2},\ldots, \bold{X_P}\)), the parameters ($\theta_0$ and $\phi_0$) of the two mapping functions ($f_\theta$ and $h_\phi$) are meta-initialized. To do so, based on the randomly sampled projection set ($\bold{X}$), the two optimization loops (i.e., inner loop and outer loop) are iterated sequentially, resulting in the final parameters of $\theta_0^*$ and $\phi_0^*$. Here, $\gamma$ is an inner loop learning rate, and $\epsilon$ is an outer loop learning rate.

Fig.~\ref{fig:1} a. illustrates the reconstruction process of the FACT method with (solid line in Fig.~\ref{fig:1} a) and without (dotted line in Fig.~\ref{fig:1} a) the meta-initialization. With the meta-initialization, since the mapping functions are already pre-trained to learn CBCT prior, the FACT reconstruction quickly and accurately optimizes each target scan to a specific CBCT image. Furthermore, it requires no additional computational cost at the reconstruction time.

%%%%%%%%%%%%%%%%%%%%%%%%%%%%%%%%%%%%%%%%%%%%%%%%%%%%%%%%%%%%%%%%%%%%%%%%%%%%%%%%%%%%%%%%%%%%%%%%%%%%%%%%%%%%%%%%%%%%%%%%%%%%%%%%

\begin{table*}[t]
\caption{\label{tab1}3D SSIM and 3D PSNR values depending on the different numbers of input views for the CBCT reconstruction methods}
\centering
\begin{tabular}{P{80pt}P{25pt}P{25pt}P{25pt}P{25pt}P{25pt}P{25pt}P{25pt}P{25pt}P{25pt}P{25pt}}
\toprule[1.5pt]
% Number of rows ($s$) & Epochs between \newline resolutions ($T$) & PSNR & SSIM \\

\multirow{2}{*}{\begin{tabular}[c]{@{}c@{}}Optimization \\ method\end{tabular}} & 
\multicolumn{2}{c}{50 Views} & 
\multicolumn{2}{c}{40 Views} & 
\multicolumn{2}{c}{30 Views} & 
\multicolumn{2}{c}{20 Views} & 
\multicolumn{2}{c}{10 Views} \\
\cline{2-11}

& PSNR & SSIM & PSNR & SSIM & PSNR & SSIM & PSNR & SSIM & PSNR & SSIM \\

\hline

FDK & 18.6 & 0.48 & 17.5 & 0.43 & 16.9 & 0.37 & 16.0 & 0.29  & 14.1 & 0.18 \\
SART & 27.2 & 0.72 & 27.5 & 0.72 & 26.7 & 0.68 & 25.5 & 0.62  & 22.9 & 0.51 \\
ASD-POCS & 29.1 & 0.79 & 28.7 & 0.77 & 28.0 & 0.74 & 26.8 & 0.69  & 23.7 & 0.56 \\
NAF & 29.9 & 0.81 & 28.9 & 0.77 & 28.0 & 0.73 & 27.0 & 0.69  & 23.6 & 0.59 \\
FACT (w/o meta) & 30.9 & 0.83 & 30.6 & 0.82 & 29.8 & 0.79 & 28.2 & 0.73  & 25.5 & 0.65 \\
FACT (w/o reg) & 31.6 & 0.84 & 31.2 & 0.83 & 30.4 & 0.80 & 29.1 & 0.75  & 26.7 & 0.67 \\
FACT (Ours) & \textbf{32.0} & \textbf{0.85} & \textbf{31.5} & \textbf{0.84} & \textbf{30.7} & \textbf{0.81} & \textbf{29.5} & \textbf{0.77}  & \textbf{27.0} & \textbf{0.70} \\

\bottomrule[1.5pt]
\end{tabular}
\end{table*}

%%%%%%%%%%%%%%%%%%%%%%%%%%%%%%%%%%%%%%%%%%%%%%%%%%%%%%%%%%%%%%%%%%%%%%%%%%%%%%%%%%%%%%%%%%%%%%%%%%%%%%%%%%%%%%%%%%%%%%%%%%%%%%%%

\subsubsection{FACT Reconstruction with Hash-Encoding Regularization}\label{section:3.2.2}
In the hash-encoding process \citep{muller2022instant}, 3D coordinates of points along the trajectory of a virtual ray were encoded using the hash-encoder (see dotted line in Fig.~\ref{fig:1} b and $h_\phi$ in Section \ref{section:3.1}). The hash-encoder is a lookup table that includes feature vectors of multi-resolution 3D grid points. In each resolution, learnable feature vectors were assigned in the corner points (e.g., corner points of the yellow or green box in Fig.~\ref{fig:1} b). When a query point with a 3D coordinate is given (e.g., $p_2$ in Fig.~\ref{fig:1} b), the feature vectors of neighboring corner points were tri-linearly interpolated to generate new feature embeddings of the query point for each resolution (e.g., interpolated features for the $1^{st}$ and $2^{nd}$ coarsest resolutions in Fig.~\ref{fig:1} b). Then these interpolated features were concatenated to a single vector (i.e., hash-encoded feature) and fed into the neural network ($f_\theta$ in Section \ref{section:3.1}) to produce an attenuation coefficient (e.g., $\mu_2$ in Fig.~\ref{fig:1} b).

In our FACT method, a new way to regularize the hash-encoder is introduced by masking out the hash-encoded feature depending on the optimization epochs (see dotted line vs. solid line in Fig.~\ref{fig:1} b). Given a hash-encoded feature vector $h_\phi$ ($\in R^{L\times F}$, where $L$ is the number of resolution levels, and $F$ is the number of feature dimensions per entry; see Table 1 in \cite{muller2022instant} for notations), we multiplied a binary mask $m$ ($\in R^{L\times F}$) to $h_\phi(p)$, so it can regularize the visible resolution through the optimization as follows:

\begin{equation}
\hat{h}_\phi(t,s;p)=h_\phi(p) \odot m(t,s)
\end{equation}

$$
m_i(t,s)=
\begin{cases}
1 & \text{if } i \leq r\\
0 & \text{if } i > r
\end{cases}
\quad \text{where } r = \left\lfloor \frac{t}{T} \right\rfloor + s 
$$

where $\hat{h}_\phi$ is the masked hash-encoded feature (see Fig.~\ref{fig:1} b) inputted to the neural network, $m_i (t,s)$ denotes the $i^{th}$ row of the $m(t,s)$, $t$ is a current epoch, $T$ is an incremental number of epochs between resolutions, and $s$ determines the number of rows of the masked hash-encoded feature to be used in an initial epoch.

The red box in Fig.~\ref{fig:1} b shows the case when the parameters are $L$ = 3, $F$ = 2, $s$ = 1, and $T$ = 100 as a simple example. In this case, until 100 epochs, only the interpolated feature of the $1^{st}$ coarsest resolution is considered as an input to the neural network, while the other interpolated features are masked. Then, after 100 epochs, the interpolated feature of the $2^{nd}$ coarsest resolution is additionally considered, and so on. This procedure regularizes the resolution levels of an input to the network to prevent high-resolution overfitting at the start of the optimization, similar to \cite{yang2023freenerf}.

\subsection{Data Preparation and Implementation Details} \label{section:3.3}
We collected publicly available human organ CT datasets of the chest, abdomen, and head. The CT scans were from the LIDC-IDRI dataset for the chest \citep{armato2011lung}, the AbdomenCT-1K dataset for the abdomen \citep{ma2021abdomenct}, and the Head-Neck-PET-CT dataset for the head \citep{vallieres2017radiomics}. For each dataset, we randomly sampled 15 CT scans for meta-learning and 15 CT scans for evaluation. The chest CT dataset provided detailed information about CT vendors for each scan, and, for meta-learning, we equally distributed 15 CT scans based on the three most prevalent vendors (Siemens, GE, and Phillips; 5 scans for each vendor). As a pre-processing step, all the CT images were first resampled to have isotropic voxels ($1\;mm^3$) and reshaped to a size of $256 \times 256 \times 256$.

To acquire X-ray projections using a cone-beam source, we generated digitally reconstructed radiographs based on a CBCT simulator (tomographic iterative GPU-based reconstruction toolbox; TIGRE) \citep{biguri2016tigre}. The TIGRE tool repetitively simulated different numbers of X-rays five times (50, 40, 30, 20, and 10 X-rays) for each CT scan with equally distributed views between $0^\circ$ and $180^\circ$. The parameters we used for the simulation were as follows: distance source-detector = 2000, distance source origin = 1000, detector resolution = (256, 256), detector pixel size = (2.0, 2.0), CT resolution = (256, 256, 256), and CT voxel size = (1.0, 1.0, 1.0).

The neural network's architecture to output an attenuation coefficient (see Fig.~\ref{fig:1} b) had four layers with 32 hidden nodes. For the meta-learning, we set the number of epochs ($K$) = 200, an inner loop learning rate ($\gamma$) = 0.001, and an outer loop learning rate ($\epsilon$) = 0.001 (see Algorithm \ref{algorithm:1} in Section \ref{section:3.2.1}). For the parameters of the hash-encoder, we set the number of levels ($L$) = 8, number of feature dimensions per entry ($F$) = 2, incremental number of epochs between resolutions ($T$) = 25, number of rows of the masked hash-encoded feature to be used in an initial epoch ($s$) = 3, coarsest resolution ($N_{min}$) = 8, and hash table size ($T$) = $2^{19}$ (see Section~\ref{section:3.2.2} in details and Table 1 in \cite{muller2022instant} for notations). The other parameters include a batch size = 256 and the maximum number of epochs for optimization at test time = 1,500.

Our hardware includes GPU = NVIDIA RTX 3090, CPU = Intel(R) Xeon(R) Silver 4310, and memory = 64 GB. We used Python 3.9.16 and PyTorch 1.11.0 for implementation.

%%%%%%%%%%%%%%%%%%%%%%%%%%%%%%%%%%%%%%%%%%%%%%%%%%%%%%%%%%%%%%%%%%%%%%%%%%%%%%%%%%%%%%%%%%%%%%%%%%%%%%%%%%%%%%%%%%%%%%%%%%%%%%%%

\begin{figure}[!th]
\centering
\includegraphics[scale=.5]{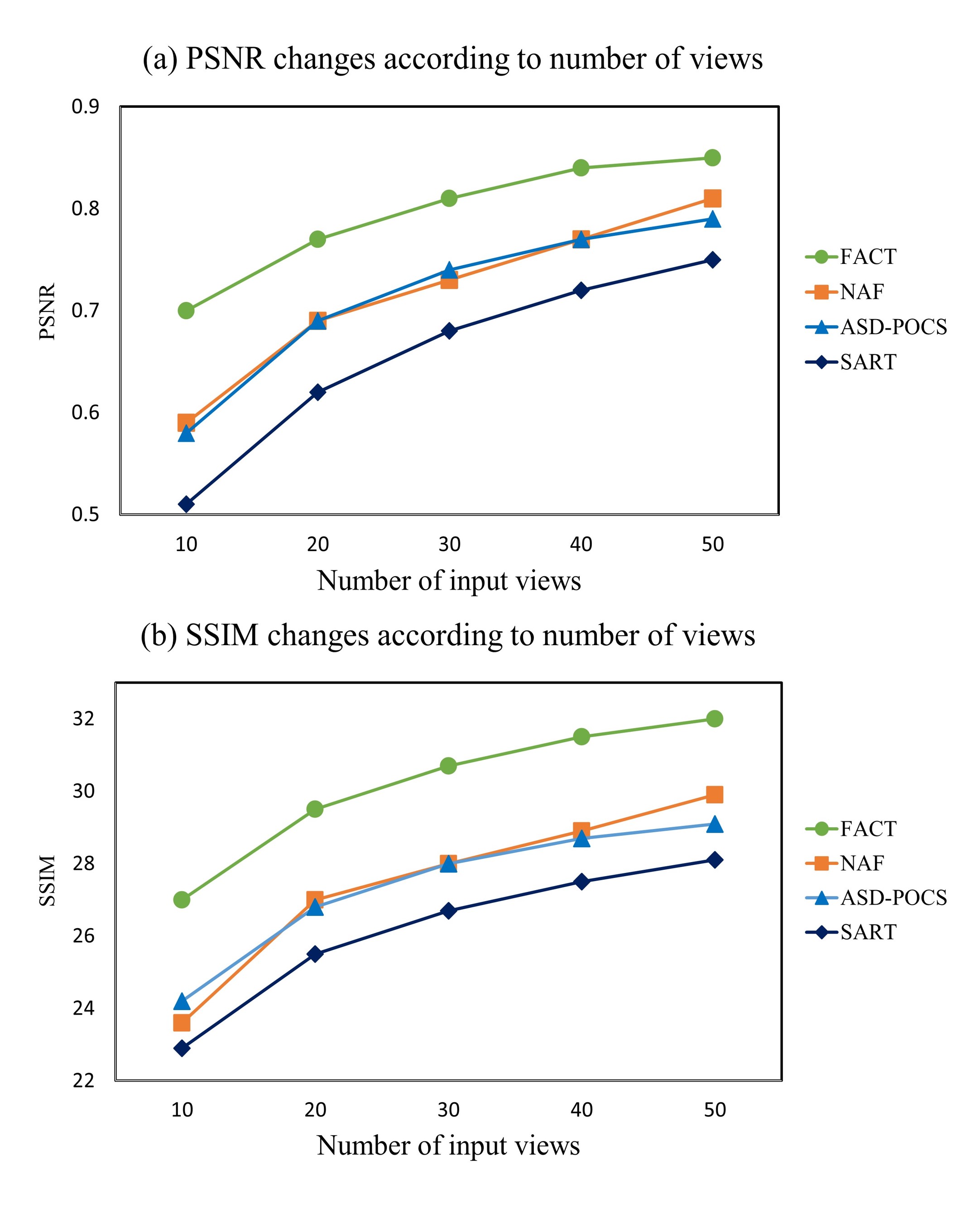}
\caption{3D SSIM and 3D PSNR values according to the change in the number of input views (50, 40, 30, 20, and 10 views) for the SART, ASD-POCS, NAF, and FACT methods. Regardless of the number of views, the FACT method outperformed the others regarding SSIM and PSNR. Especially in the graph of SSIM, FACT revealed more gains as the number of views decreased.}
\label{fig:2}
\end{figure}

%%%%%%%%%%%%%%%%%%%%%%%%%%%%%%%%%%%%%%%%%%%%%%%%%%%%%%%%%%%%%%%%%%%%%%%%%%%%%%%%%%%%%%%%%%%%%%%%%%%%%%%%%%%%%%%%%%%%%%%%%%%%%%%%

%%%%%%%%%%%%%%%%%%%%%%%%%%%%%%%%%%%%%%%%%%%%%%%%%%%%%%%%%%%%%%%%%%%%%%%%%%%%%%%%%%%%%%%%%%%%%%%%%%%%%%%%%%%%%%%%%%%%%%%%%%%%%%%%

\begin{figure*}[!th]
\centering
\includegraphics[scale=0.9]{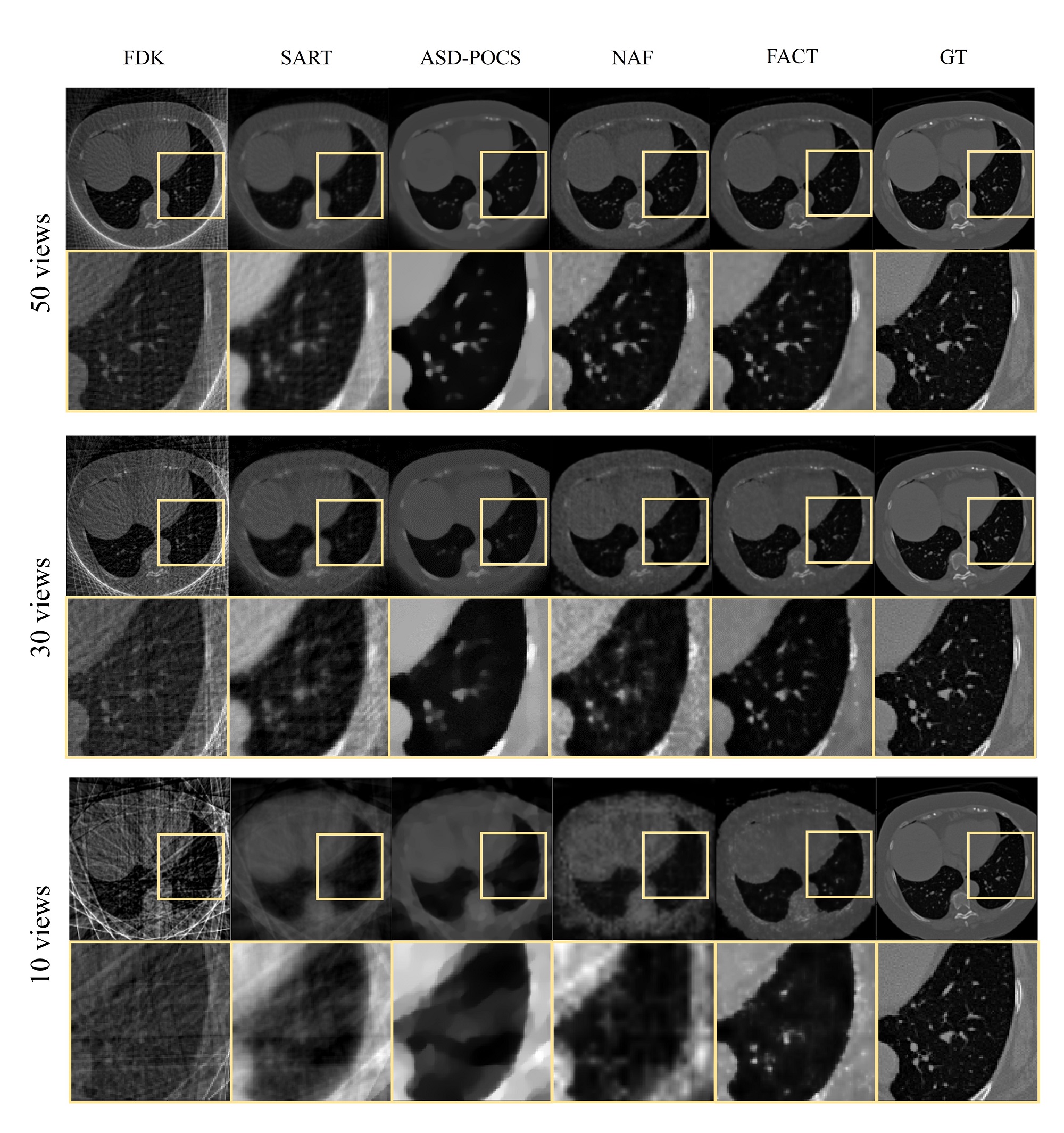}
\caption{Example CBCT reconstruction results between the different CBCT reconstruction methods (FDK, SART, ASD-POCS, NAF, and FACT). In the zoomed-in plots (yellow boxes), the FACT method demonstrated better image quality, successfully reconstructing anatomical structures and suppressing artifacts than the other methods (e.g., results of NAF vs. FACT in 30 views).}
\label{fig:3}
\end{figure*}

%%%%%%%%%%%%%%%%%%%%%%%%%%%%%%%%%%%%%%%%%%%%%%%%%%%%%%%%%%%%%%%%%%%%%%%%%%%%%%%%%%%%%%%%%%%%%%%%%%%%%%%%%%%%%%%%%%%%%%%%%%%%%%%%

\subsection{Experiments}
We compared the quality of the CBCT reconstruction results using two metrics, 3D structural similarity index measure (SSIM) and 3D peak signal-to-noise ratio (PSNR), between the FDK, SART, ASD-POCS, NAF, and FACT methods for each 50, 40, 30, 20, and 10 views. The average values of those metrics were calculated over the 15 testing images for each human organ (chest, head, and abdomen). Also, to check the consistency of the reconstruction results, we reported the 3D SSIM and 3D PSNR values depending on the different CT vendors (Siemens, Phillips, and GE; five testing chest CTs for each) between the NAF and FACT methods.

To compare the optimization speed between all the methods (FDK, SART, ASD-POCS, NAF, and FACT), we measured the mean running times (in minutes) of each method over the 15 testing images of the chest. To measure the running times of the conventional algorithms (FDK, SART, and ASD-POCS), we used pre-implemented functions of the FDK, SART (50 iterations), and ASD-POCS (20 iterations with 25 total variation minimization steps) in the TIGRE library. For the running times of the NAF and FACT methods, we calculated the time when the variation of 3D PSNR values was converged within 0.01 for 50 consecutive epochs (i.e., convergence time). To run the NAF algorithm, we utilized a source code from \cite{zha2022naf} on the web and followed the same parameters in Section \ref{section:3.3}.

For the key parameters of the hash-encoding regularization, including the incremental number of epochs between resolutions ($T$) and number of rows of the masked hash-encoded feature to be used in an initial epoch ($s$), we performed a manual search to find the optimal parameters (see Section~\ref{section:4.3}).

%%%%%%%%%%%%%%%%%%%%%%%%%%%%%%%%%%%%%%%%%%%%%%%%%%%%%%%%%%%%%%%%%%%%%%%%%%%%%%%%%%%%%%%%%%%%%%%%%%%%%%%%%%%%%%%%%%%%%%%%%%%%%%%%

\begin{table*}[h]
\caption{\label{tab2}3D PSNR and 3D SSIM values for the CBCT reconstruction results of the head and abdomen}
\centering
\begin{tabular}{P{50pt}P{60pt}P{25pt}P{25pt}P{25pt}P{25pt}P{25pt}P{25pt}P{25pt}P{25pt}P{25pt}P{25pt}}
% \begin{tabular}{P{65pt}P{65pt}P{35pt}P{5pt}P{35pt}P{5pt}P{35pt}P{35pt}}
\toprule[1.5pt]
% Number of rows ($s$) & Epochs between \newline resolutions ($T$) & PSNR & SSIM \\

\multirow{2}{*}{Body part} & 
\multirow{2}{*}{\begin{tabular}[c]{@{}c@{}}Optimization \\ method\end{tabular}} & 
\multicolumn{2}{c}{50 Views} & 
\multicolumn{2}{c}{40 Views} & 
\multicolumn{2}{c}{30 Views} & 
\multicolumn{2}{c}{20 Views} & 
\multicolumn{2}{c}{10 Views} \\
\cline{3-12}

& & PSNR & SSIM & PSNR & SSIM & PSNR & SSIM & PSNR & SSIM & PSNR & SSIM \\

\hline

\multirow{5}{*}{Head} & FDK & 20.1 & 0.45 & 20.7 & 0.40 & 19.8 & 0.34 & 18.6 & 0.26  & 15.9 & 0.16 \\

& SART & 29.7 & 0.83 & 29.3 & 0.82 & 28.4 & 0.78 & 26.9 & 0.71  & 24.1 & 0.57 \\
& ASD-POCS & 31.1 & 0.89 & 30.5 & 0.87 & 29.0 & 0.84 & 28.0 & 0.78  & 25.6 & 0.68 \\
& NAF & 32.3 & 0.90 & 31.6 & 0.89 & 30.4 & 0.86 & 28.7 & 0.79  & 26.2 & 0.72 \\
& FACT (Ours) & \textbf{33.8} & \textbf{0.92} & \textbf{32.9} & \textbf{0.91} & \textbf{31.9} & \textbf{0.89} & \textbf{30.3} & \textbf{0.84}  & \textbf{27.9} & \textbf{0.77} \\

\hline

\multirow{5}{*}{Abdomen} & FDK & 16.5 & 0.49 & 18.5 & 0.48 & 17.9 & 0.43 & 17.0 & 0.36 & 13.8 & 0.21 \\

& SART & 28.1 & 0.81 & 28.3 & 0.80 & 27.6 & 0.77 & 26.4 & 0.72  & 23.4 & 0.61 \\
& ASD-POCS & 30.3 & 0.87 & 29.7 & 0.86 & 28.9 & 0.83 & 27.6 & 0.79  & 25.3 & 0.71 \\
& NAF & 31.4 & 0.87 & 30.6 & 0.84 & 29.9 & 0.81 & 28.5 & 0.77  & 26.0 & 0.71 \\
& FACT (Ours) & \textbf{32.9} & \textbf{0.91} & \textbf{32.2} & \textbf{0.89} & \textbf{31.3} & \textbf{0.86} & \textbf{30.0} & \textbf{0.82}  & \textbf{27.2} & \textbf{0.76} \\

\bottomrule[1.5pt]
\end{tabular}
\end{table*}

%%%%%%%%%%%%%%%%%%%%%%%%%%%%%%%%%%%%%%%%%%%%%%%%%%%%%%%%%%%%%%%%%%%%%%%%%%%%%%%%%%%%%%%%%%%%%%%%%%%%%%%%%%%%%%%%%%%%%%%%%%%%%%%%

%%%%%%%%%%%%%%%%%%%%%%%%%%%%%%%%%%%%%%%%%%%%%%%%%%%%%%%%%%%%%%%%%%%%%%%%%%%%%%%%%%%%%%%%%%%%%%%%%%%%%%%%%%%%%%%%%%%%%%%%%%%%%%%%

\begin{figure*}[!t]
\centering
\includegraphics[scale=0.9]{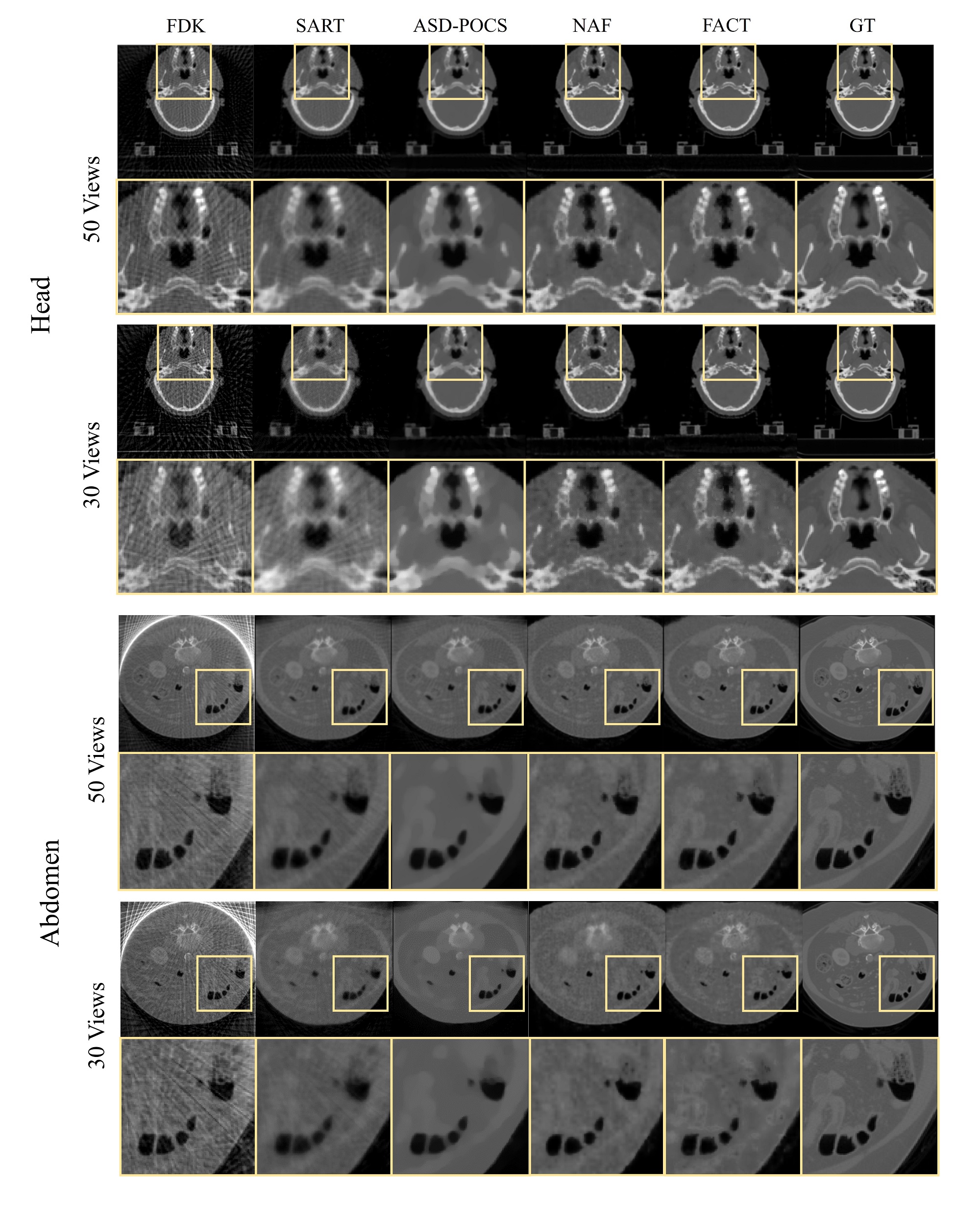}
\caption{Example head and abdomen CBCT reconstruction results, including FDK, SART, ASD-POCS, NAF, and FACT, for 50 and 30 views. The results of the FACT method showed the best quality compared to the others.}
\label{fig:4}
\end{figure*}

%%%%%%%%%%%%%%%%%%%%%%%%%%%%%%%%%%%%%%%%%%%%%%%%%%%%%%%%%%%%%%%%%%%%%%%%%%%%%%%%%%%%%%%%%%%%%%%%%%%%%%%%%%%%%%%%%%%%%%%%%%%%%%%%

%%%%%%%%%%%%%%%%%%%%%%%%%%%%%%%%%%%%%%%%%%%%%%%%%%%%%%%%%%%%%%%%%%%%%%%%%%%%%%%%%%%%%%%%%%%%%%%%%%%%%%%%%%%%%%%%%%%%%%%%%%%%%%%%

\begin{table}[h]
\centering
\caption{\label{tab3}3D SSIM and 3D PSNR values for the CBCT reconstruction results between the NAF and FACT methods according to the three CT vendors (Siemens, Phillips, and GE)}
\begin{tabular}{P{40pt}P{40pt}P{45pt}P{30pt}P{30pt}}
\toprule[1.5pt]

                          & Metric                & Vendor  & NAF  & FACT          \\ \hline
\multirow{6}{*}{50 Views} & \multirow{3}{*}{PSNR} & Siemens & 28.1 & \textbf{30.7} \\ 
                          &                       & Philips & 31.6 & \textbf{33.3} \\  
                          &                       & GE      & 30.0 & \textbf{31.9} \\ \cline{2-5} 
                          & \multirow{3}{*}{SSIM} & Siemens & 0.76 & \textbf{0.81} \\ 
                          &                       & Philips & 0.84 & \textbf{0.88} \\  
                          &                       & GE      & 0.82 & \textbf{0.86} \\ \hline
\multirow{6}{*}{30 Views} & \multirow{3}{*}{PSNR} & Siemens & 26.2 & \textbf{29.5} \\ 
                          &                       & Philips & 29.6 & \textbf{32.1} \\  
                          &                       & GE      & 28.4 & \textbf{30.5} \\ \cline{2-5} 
                          & \multirow{3}{*}{SSIM} & Siemens & 0.66 & \textbf{0.77} \\ 
                          &                       & Philips & 0.77 & \textbf{0.85} \\ 
                          &                       & GE      & 0.76 & \textbf{0.82} \\ \hline
\multirow{6}{*}{10 Views} & \multirow{3}{*}{PSNR} & Siemens & 21.4 & \textbf{29.1} \\ 
                          &                       & Philips & 25.7 & \textbf{31.7} \\ 
                          &                       & GE      & 23.6 & \textbf{30.0} \\ \cline{2-5} 
                          & \multirow{3}{*}{SSIM} & Siemens & 0.51 & \textbf{0.75} \\ 
                          &                       & Philips & 0.65 & \textbf{0.83} \\ 
                          &                       & GE      & 0.60 & \textbf{0.80} \\ 

\bottomrule[1.5pt]
\end{tabular}
\end{table}

%%%%%%%%%%%%%%%%%%%%%%%%%%%%%%%%%%%%%%%%%%%%%%%%%%%%%%%%%%%%%%%%%%%%%%%%%%%%%%%%%%%%%%%%%%%%%%%%%%%%%%%%%%%%%%%%%%%%%%%%%%%%%%%%

\section{Results}
\subsection{CBCT Reconstruction Results}
Fig.~\ref{fig:2} shows the graphs that report 3D SSIM and 3D PSNR values according to the change in the number of input views (50, 40, 30, 20, and 10 views) for the SART, ASD-POCS, NAF, and FACT methods. Regardless of the number of views, the FACT method outperformed the others (see Table~\ref{tab1} for details, including FDK). Especially in the graph of SSIM (Fig.~\ref{fig:2} a), FACT revealed more gains as the number of views decreased compared to the others. When the number of views was very limited (e.g., 30 views). Indeed, when we investigated the reconstructed images (Fig.~\ref{fig:3}), the FACT method showed better image quality, disclosing the anatomical structures and suppressing artifacts than the other methods (see zoomed-in yellow boxes in Fig.~\ref{fig:3}).

Table~\ref{tab1} also reports the FACT method's 3D PSNR and 3D SSIM values, excluding the meta-initialization or hash-encoding regularization. When we ablated the hash-encoding regularization (i.e., FACT w/o reg. in Table~\ref{tab1}), the reconstruction quality was degraded (e.g., 30.7 vs. 30.4 in 3D PSNR for 30 views). However, ablating the meta-initialization showed a more significant performance drop (e.g., 30.7 and 29.8 in 3D PSNR for 30 views). The best performance was achieved when the meta-initialization and hash-encoding regularization were combined.

Table~\ref{tab2} summarizes 3D PSNR and 3D SSIM values for the CBCT reconstruction results of different body parts (head and abdomen). FACT showed the highest scores in terms of PSNR and SSIM. Fig.~\ref{fig:4} displays the head and abdomen examples for 50 and 30 views, and the results of FACT demonstrated the best image quality, which was in line with the performance metrics in Table~\ref{tab2}.

Table~\ref{tab3} summarizes 3D SSIM and 3D PSNR values of the CBCT reconstruction results between the NAF and FACT methods according to the three CT vendors (Siemens, Phillips, and GE) for the chest CBCT. Regardless of the CT vendors, the FACT method consistently outperformed the NAF method (e.g., 3D PSNR for 30 views: 26.2 vs. 29.5 for Siemens, 29.6 vs. 32.1 for Phillips, and 28.4 vs. 30.5 for GE), implying that FACT did not have a substantial bias toward a specific CT vendor after the meta-initialization. Additionally, Supplementary Fig. 1 shows example reconstruction results for 30 views.

%%%%%%%%%%%%%%%%%%%%%%%%%%%%%%%%%%%%%%%%%%%%%%%%%%%%%%%%%%%%%%%%%%%%%%%%%%%%%%%%%%%%%%%%%%%%%%%%%%%%%%

\begin{table*}[h]
\caption{\label{tab4}Number of epochs and optimization times of the NAF and FACT methods to reach the best performance of the NAF method}
\centering
\begin{tabular}{P{65pt}P{65pt}P{50pt}P{40pt}P{50pt}P{40pt}}
% \begin{tabular}{P{65pt}P{65pt}P{35pt}P{5pt}P{35pt}P{5pt}P{35pt}P{35pt}}
\toprule[1.5pt]
% Number of rows ($s$) & Epochs between \newline resolutions ($T$) & PSNR & SSIM \\

\multirow{2}{*}{Number of views} & 
\multirow{2}{*}{\begin{tabular}[c]{@{}c@{}}Optimization \\ method\end{tabular}} & 
\multicolumn{2}{c}{\begin{tabular}[c]{@{}c@{}}Number of epochs to reach\\ best NAF performance \end{tabular}} & 
\multicolumn{2}{c}{\begin{tabular}[c]{@{}c@{}}Optimization time to reach\\ best NAF performance (sec.)\end{tabular}} \\
\cline{3-6}

& & 3D PSNR & 3D SSIM & 3D PSNR & 3D SSIM \\

\hline

\multirow{2}{*}{50 Views} & NAF & 1490 & 1467 & 677 & 667 \\

& FACT & \textbf{256} & \textbf{303} & \textbf{116} & \textbf{138} \\

\hline

\multirow{2}{*}{30 Views} & NAF & 1363 & 1456 & 371 & 397 \\

& FACT & \textbf{172} & \textbf{197} & \textbf{47} & \textbf{54} \\

\hline

\multirow{2}{*}{10 Views} & NAF & 1380 & 1420 & 125 & 129 \\

& FACT & \textbf{145} & \textbf{103} & \textbf{13} & \textbf{9} \\

\bottomrule[1.5pt]
\end{tabular}
\end{table*}

%%%%%%%%%%%%%%%%%%%%%%%%%%%%%%%%%%%%%%%%%%%%%%%%%%%%%%%%%%%%%%%%%%%%%%%%%%%%%%%%%%%%%%%%%%%%%%%%%%%%%%%%%%%%%%%%%%%%%%%%%%%%%%%%

\subsection{Optimization Speed of CBCT Reconstruction}
\begin{figure}[!t]
\centering
\includegraphics[scale=.4]{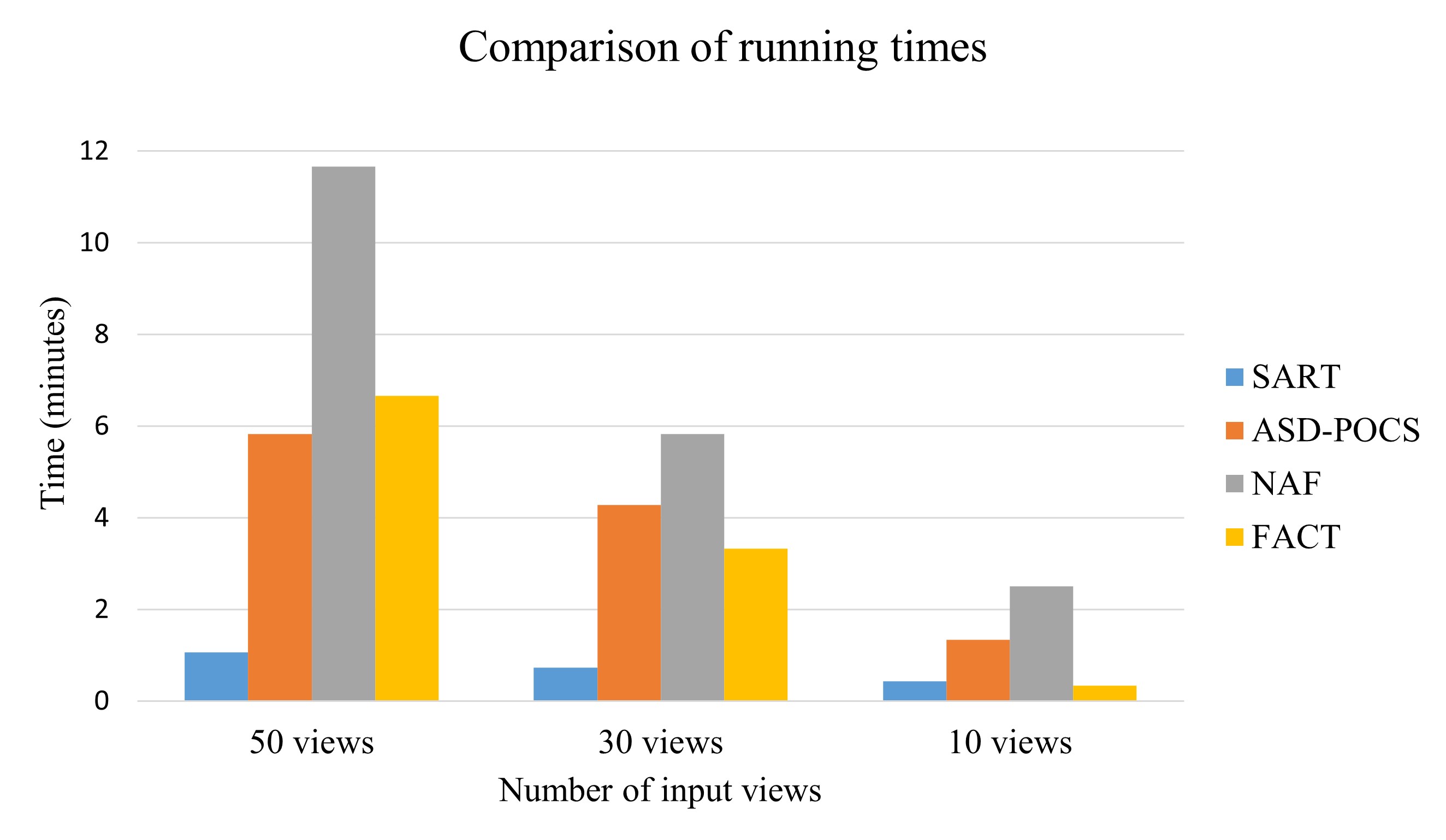}
\caption{Comparison of the running times of the different reconstruction methods (SART, ASD-POCS, NAF, and FACT). For some cases, FACT showed comparable running times to those of the conventional methods (SART and ASD-POCS). For example, in the case of 30 views, the running time of FACT was faster than that of ASD-POCS but slower than that of SART. However, in the case of 10 views, the running time of FACT was even slightly faster than SART.}
\label{fig:5}
\end{figure}

Fig.~\ref{fig:5} compares the running times of the different reconstruction methods (SART, ASD-POCS, NAF, and FACT). The running times of the conventional algorithms (FDK, SART, and ASD-POCS; FDK took a few seconds for all cases) were relatively fast compared to those of the NAF and FACT methods. In all the configurations, NAF required the longest time for reconstruction, between 2 minutes and 12 minutes, while FACT showed comparable running times to those of the conventional methods for some cases. For example, in the case of 30 views, the running time of FACT was faster than that of ASD-POCS but slower than that of SART. However, in the case of 10 views, the running time of FACT was even slightly faster than SART.

Table~\ref{tab4} summarizes the number of epochs and optimization times of the NAF and FACT methods to reach the best performance of the NAF method. In the case of 30 views, the FACT method reached the best performance of NAF, spending less than 14 percent of the number of epochs and optimization times (e.g., 1456 epochs and 397 seconds for NAF vs. 197 epochs and 54 seconds for FACT in 3D SSIM). The gap was more significant in the case of 10 views where the FACT method only required less than 8 percent of the number of epochs and optimization times than the NAF method to reach the same performance (e.g., 1420 epochs and 129 seconds for NAF vs.103 epochs and 9 seconds for FACT in 3D SSIM).

\begin{figure}[!t]
\centering
\includegraphics[scale=.4]{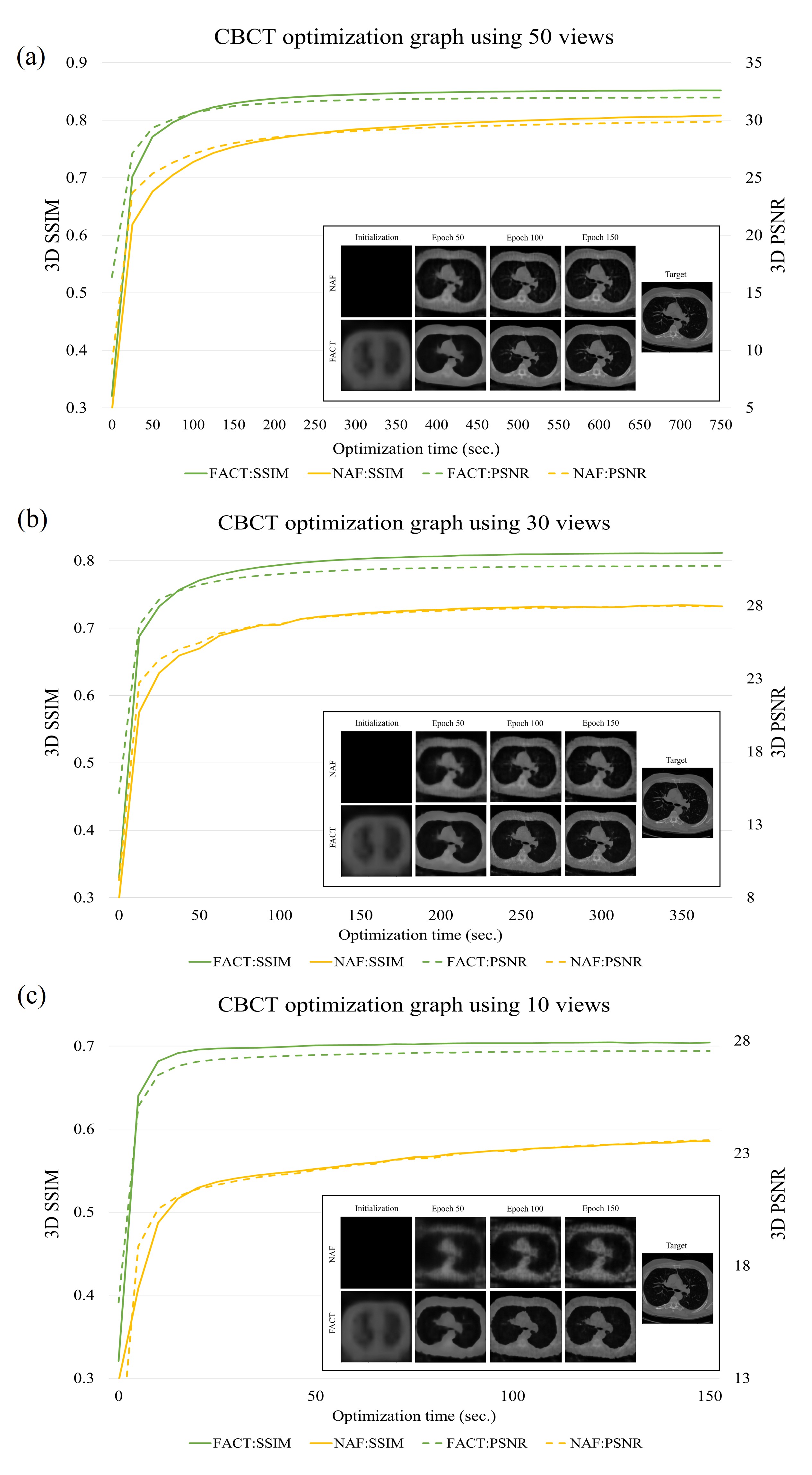}
\caption{CBCT optimization graphs (3D SSIM or 3D PSNR vs. optimization epochs) between the NAF and FACT methods for 50, 30, and 10 views. The graphs of the FACT method consistently are above the graphs of the NAF method, demonstrating faster optimization speed and better reconstruction quality. The intermediate reconstruction results of the FACT method are shown below in each graph. At the initial optimization stage, the NAF method produced an empty image (i.e., no information), while the FACT method generated a CT-like image.}
\label{fig:6}
\end{figure}

Fig.~\ref{fig:6}. shows 3D SSIM and 3D PSNR values according to the change of the optimization epochs between the NAF and FACT methods for 50, 30, and 10 views (see Supplementary Fig. 2 for the graphs of the head and abdomen). In each graph, sample reconstruction results are plotted at initialization, 50 epochs, 100 epochs, and 150 epochs. The optimization graphs of the FACT method consistently are above the graphs of the NAF method, demonstrating faster optimization speed and better reconstruction quality. When we investigated the intermediate optimization results in detail, especially in the graph of 10 views (Fig.~\ref{fig:2} c), the NAF method produced unsatisfactory results, failing to reconstruct the delicate structures of the CBCT images compared to the FACT method. At the initial optimization stage, the NAF method provided an empty image (i.e., no information), while the FACT method, which was meta-trained with a few CBCTs, still generated a CT-like image.

%%%%%%%%%%%%%%%%%%%%%%%%%%%%%%%%%%%%%%%%%%%%%%%%%%%%%%%%%%%%%%%%%%%%%%%%%%%%%%%%%%%%%%%%%%%%%%%%%%%%%%%%%%%%%%%%%%%%%%%%%%%%%%%%

%%%%%%%%%%%%%%%%%%%%%%%%%%%%%%%%%%%%%%%%%%%%%%%%%%%%%%%%%%%%%%%%%%%%%%%%%%%%%%%%%%%%%%%%%%%%%%%%%%%%%%%%%%%%%%%%%%%%%%%%%%%%%%%%

%%%%%%%%%%%%%%%%%%%%%%%%%%%%%%%%%%%%%%%%%%%%%%%%%%%%%%%%%%%%%%%%%%%%%%%%%%%%%%%%%%%%%%%%%%%%%%%%%%%%%%%%%%%%%%%%%%%%%%%%%%%%%%%%

\begin{table}[h]
\caption{\label{tab5}Key parameter search results for the hash-encoding regularization}
\centering
\begin{tabular}{P{65pt}P{70pt}P{30pt}P{30pt}}
\toprule[1.5pt]
Number of rows ($s$) & Epochs between \newline resolutions ($T$) & \multirow{2}{*}{PSNR} & \multirow{2}{*}{SSIM} \\

\hline
2 & 100 & 30.29 & 0.788 \\

2 & 200 & 30.25 & 0.786 \\

3 & 10 & 30.38 & 0.795 \\

3 & 25 & \textbf{30.38} & 0.796 \\

3 & 50 & 30.36 & 0.796 \\

3 & 200 & 30.26 & 0.794 \\

4 & 100 & 30.29 & \textbf{0.797} \\

\bottomrule[1.5pt]
\end{tabular}
\end{table}

%%%%%%%%%%%%%%%%%%%%%%%%%%%%%%%%%%%%%%%%%%%%%%%%%%%%%%%%%%%%%%%%%%%%%%%%%%%%%%%%%%%%%%%%%%%%%%%%%%%%%%%%%%%%%%%%%%%%%%%%%%%%%%%%

\subsection{Key Parameter Search for Hash-Encoding Regularization}\label{section:4.3}
For the key parameter search of the hash-encoding regularization (see Table~\ref{tab5}), the best performance was achieved in terms of 3D PSNR when the incremental number of epochs between resolutions ($T$) was 25, and the number of rows of the masked hash-encoded feature to be used in an initial epoch ($s$) was three. While the other combination reported the best performance in 3D SSIM (i.e., case when $T$ = 100 and $s$ = 4), PSNR was largely degraded (30.38 vs. 30.29), so we decided to choose to the case when $T$ = 25 and $s$ = 3.

\section{Discussion}
The proposed FACT reconstruction method utilizes a meta-learning framework to pre-train a neural network using a small set of CBCT scans and regularizes the hash-encoding process to increase the optimization speed and the quality of the CBCT reconstruction with no additional computational cost. We proved that the 3D CBCT images of FACT had better quality than those of the other conventional algorithms. Furthermore, the FACT reconstruction speed was much faster than that of the NAF method, requiring only 14 percent (for 25 views) and 8 percent (for 10 views) of the running time to reach the same image quality. Finally, we validated that the proposed FACT method successfully reconstructed CBCTs of the multiple CT vendors (Siemens, Phillips, and GE) and body parts (chest, head, and abdomen).

As a further investigation, we also observed the intermediate CBCT reconstruction results of the different body parts at the very early stages of optimization (NAF vs. FACT in 5 and 10 epochs) to check the effect of the meta-initialization. Surprisingly, as shown in Supplementary Fig. 3, we found that the results of the FACT method produced coarse anatomical structures of each body part even after only 5 and 10 epochs, while the results of NAF could not generate any meaningful information. This means that the meta-initialization effectively accelerates the reconstruction process even at the very early optimization stage in each body part.

In the hash-encoding, we found that the finest resolution level significantly affected the quality of the CBCT reconstruction, which motivated us to regularize the hash-encoding process. Supplementary Fig. 4 shows an example where, in the NAF method, the reconstruction results with the finest resolution of 1,024 (i.e., when the number of levels (L) = 8 and coarsest resolution ($N_{min}$) = 8; see Section \ref{section:3.3}) reveals notable noises compared to the results with the finest resolution of 64 (i.e., $L$ = 4 and $N_{min}$ = 8). This observation is in line with the previous work of \citep{yang2023freenerf}, which explained the failure of the NeRF algorithm as the sudden overfitting of a neural network to high-frequency components such as noises.

This study has some limitations. First, we performed all the experiments using the simulated CBCT data since acquiring a set of raw projection images in the real world was difficult. We believe that the effectiveness of the proposed method can be further validated using real clinical data in the future. Second, although we showed that the proposed algorithm successfully reconstructed the CBCT images for some body parts, there are other body parts for CBCT imaging, such as teeth and jaws. These body parts may be tested when those CT data are publicly available. Third, some advanced encoding strategies have shown impressive results for the NeRF reconstruction, such as \cite{dou2023multiplicative}. These new techniques may be integrated into the proposed FACT algorithm to enhance the CBCT reconstruction results further.

%%%%%%%%%%%%%%%%%%%%%%%%%%%%%%%%%%%%%%%%%%%%%%%%%%%%%%%%%%%%%%%%%%%%%%%%%%%%%%%%%%%%%%%%%%%%%%%%%%%%%%%%%%%%%%%%%%%%%%%%%%%%%%%%

%%%%%%%%%%%%%%%%%%%%%%%%%%%%%%%%%%%%%%%%%%%%%%%%%%%%%%%%%%%%%%%%%%%%%%%%%%%%%%%%%%%%%%%%%%%%%%%%%%%%%%%%%%%%%%%%%%%%%%%%%%%%%%%%

\section{Conclusion}
In conclusion, with the meta-initialization and the new regularization process, the FACT method demonstrated faster and more accurate sparse-view CBCT reconstruction than the other conventional (FDK, SAR, and ASD-POCS) and NAF methods. The proposed method is expected to provide benefits in some clinical situations, such as patient positioning and emergency scans \citep{thilmann2006correction, jacques2021impact}, assuring the consistent reconstruction quality for the multiple CT machines (Siemens, Phillips, and GE) and anatomies (chest, head, and abdomen).

\section*{Acknowledgments}
Special thanks to the co-authors for their invaluable guidance and advice throughout this project. We are grateful for the contributions to the improvement of this article's readability and clarity.

\section*{Declaration of Interest}
The authors declare the following financial interests/personal relationships which may be considered as potential competing interests:
H.Shin, T.Kim, and D.Shin report a relationship with Radisen Co. Ltd. that includes: employment. J.Lee and S.Cho reports a relationship with Radisen Co. Ltd. that includes: consulting or advisory. If there are other authors, they declare that they have no known competing financial interests or personal relationships that could have appeared to influence the work reported in this paper.

\section*{Supplementary Material}
Supplementary material associated with this article can be found, in 
the online version.

%%Harvard
\bibliographystyle{model2-names.bst}\biboptions{authoryear}
\bibliography{refs}

%%%%%%%%%%%%%%%%%%%%%%%%%%%%%%%%%%%%%%%%%%%%%%%%%%%%%%%%%%%%%%%%%%%%%%%%%%%%%%%%%%%%%%%%%%%%%%%%%%%%%%%%%%%%%%%%%%%%%%%%%%%%%%%%
\clearpage
\renewcommand{\figurename}{Supplementary Figure}
\setcounter{figure}{0}

\begin{figure*}[h]
\centering
\includegraphics[scale=.5]{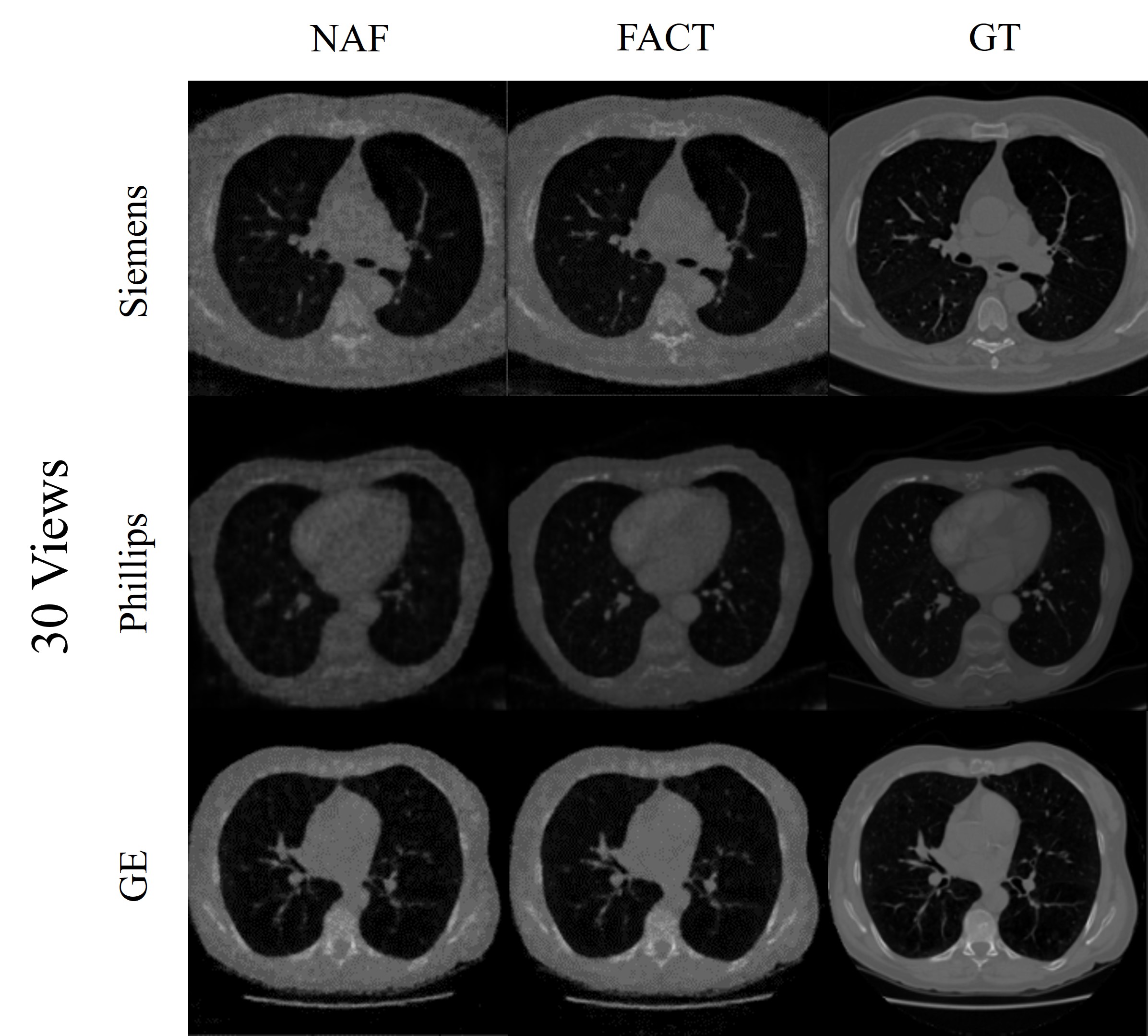}
\caption{Example CBCT reconstruction results of 30 views for the three CT vendors (Siemens, Phillips, and GE).}
\end{figure*}

\begin{figure*}[h]
\centering
\includegraphics[scale=.4]{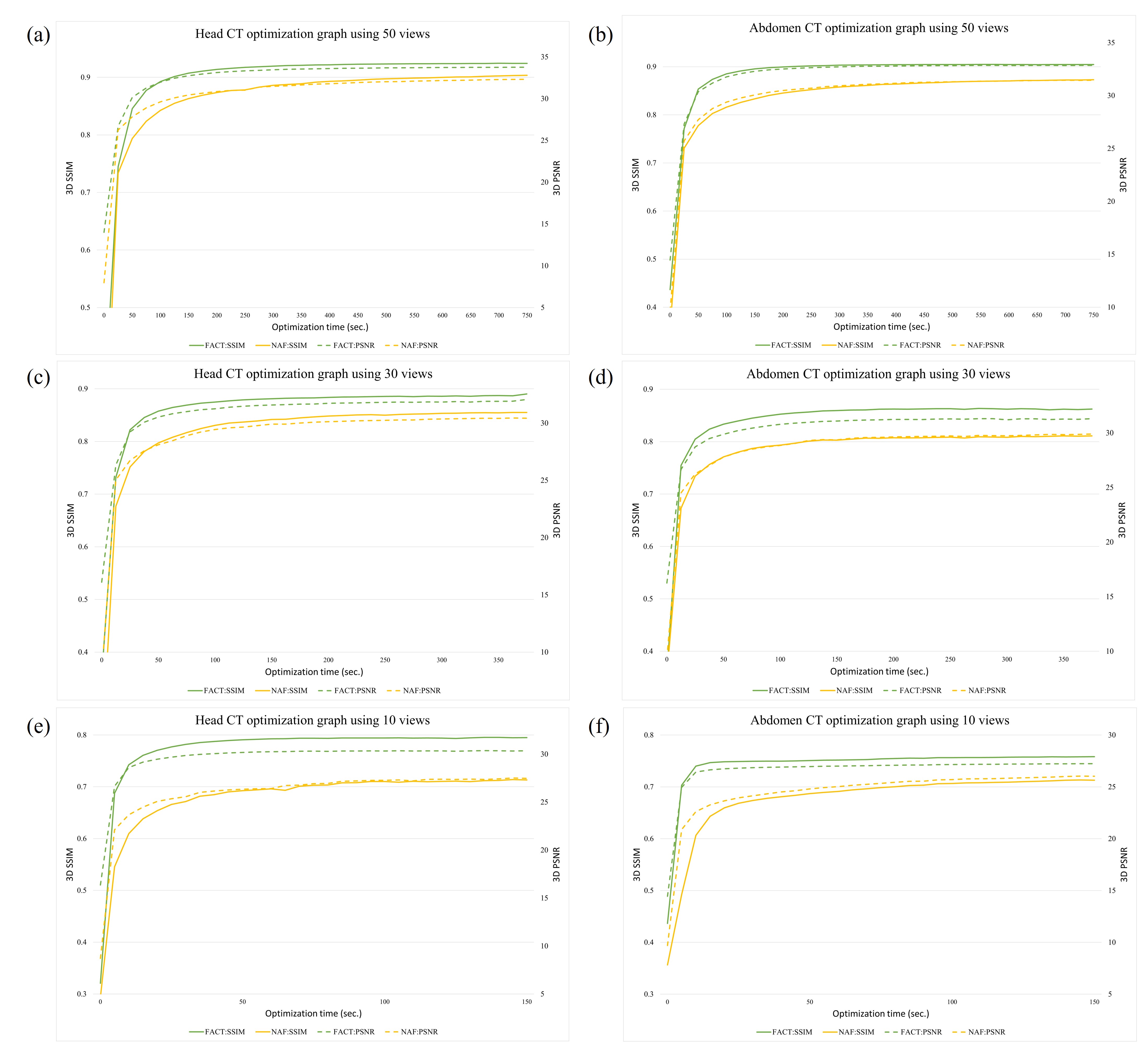}
\caption{CBCT optimization graphs (3D SSIM or 3D PSNR vs. optimization epochs) between the NAF and FACT methods for the head and abdomen CBCTs.}
\end{figure*}

\begin{figure*}[h]
\centering
\includegraphics[scale=.4]{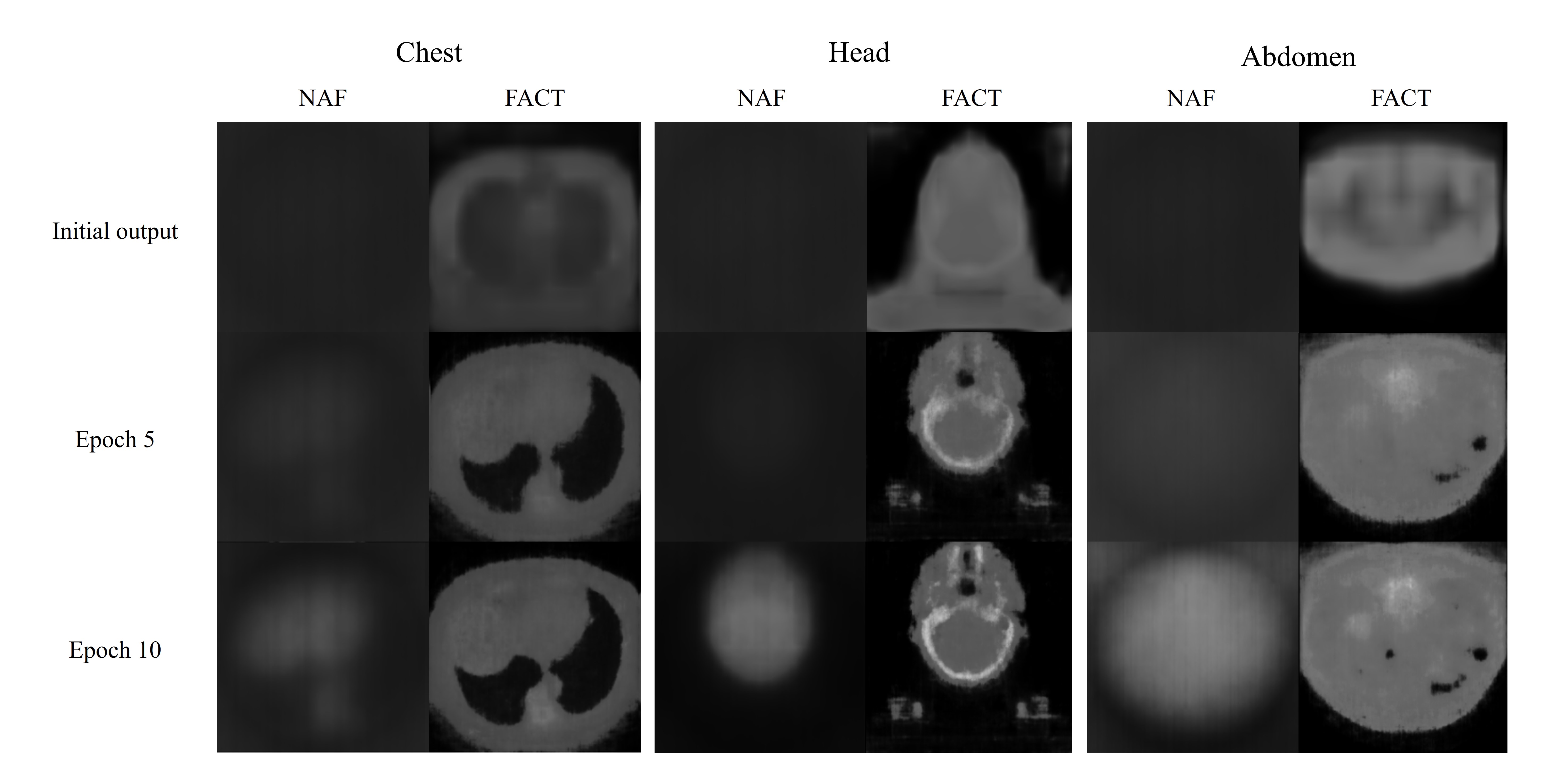}
\caption{Intermediate CBCT reconstruction results of the different body parts (chest, head, and abdomen) for the NAF and FACT methods at the very early stages of optimization (5 and 10 epochs).}
\end{figure*}

\begin{figure*}[h]
\centering
\includegraphics[scale=.8]{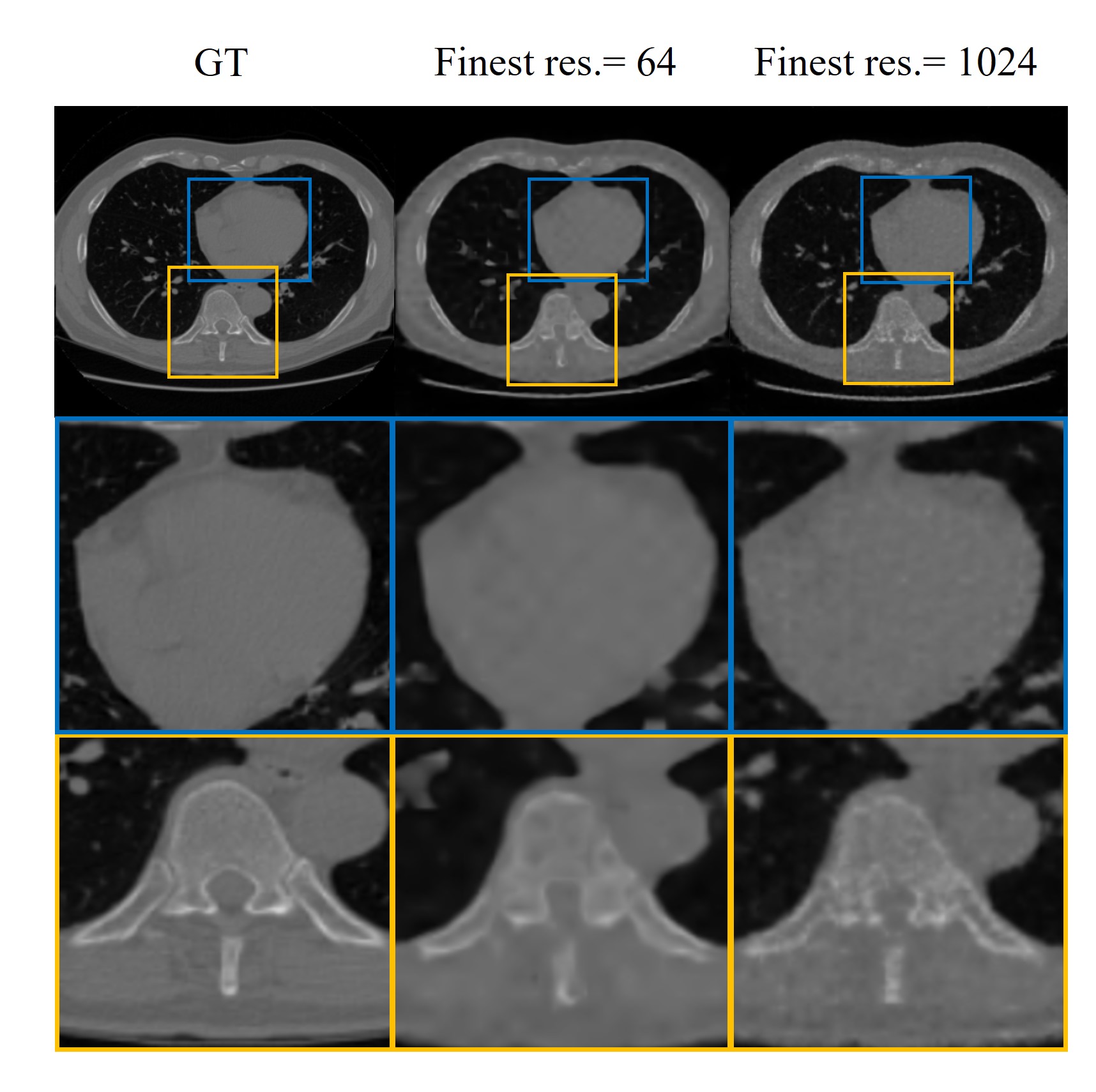}
\caption{Example CBCT reconstruction results of the NAF method when the finest resolution was 1,024 and 64. The results with the finest resolution of 1,024 reveal notable noises compared to the results with the finest resolution of 64.}
\end{figure*}

%%%%%%%%%%%%%%%%%%%%%%%%%%%%%%%%%%%%%%%%%%%%%%%%%%%%%%%%%%%%%%%%%%%%%%%%%%%%%%%%%%%%%%%%%%%%%%%%%%%%%%%%%%%%%%%%%%%%%%%%%%%%%%%%

\end{document}